\documentclass[aip,jcp,reprint,amsmath,amssymb,floatfix,citeautoscript]{revtex4-1}
\pdfoutput=1
\usepackage{cancel}
\usepackage{amsmath}
\usepackage{mathtools}
\usepackage{physics}
\usepackage{graphicx}
\usepackage{amssymb}
\usepackage{amsthm}
\usepackage{bm}
\usepackage{dcolumn}
\usepackage{braket}
\usepackage{longtable}
\usepackage{ragged2e}
\usepackage{txfonts}
\usepackage[version=3]{mhchem} 
\usepackage[T1]{fontenc}       
\usepackage{pstricks}
\usepackage{siunitx}
\usepackage[colorlinks=true,allcolors=blue]{hyperref}
\usepackage[capitalise]{cleveref}
\usepackage{multirow}
\usepackage{algorithm}         
\usepackage[noend]{algpseudocode}     
\usepackage{booktabs}
\usepackage[flushleft]{threeparttable}

\usepackage{lipsum} 

\let\tr\undefined
\newcommand*\tr[1]{\textrm{#1}}

\newcommand*\mat[1]{\mathbf{#1}}

\definecolor{blackpink}{RGB}{200, 64, 200}
\newcommand*\rvv[1]{{#1}}

\newcommand*\me{\mathrm{e}}

\newcommand*\mi{\mathrm{i}}

\allowdisplaybreaks

\begin{document}

\title
{Integral-direct Hartree-Fock and M{\o}ller-Plesset Perturbation Theory for Periodic Systems with Density Fitting: Application to the Benzene Crystal}

\author{Sylvia J. Bintrim}
\affiliation
{Department of Chemistry, Columbia University, New York, New York 10027, USA}
\author{Timothy C. Berkelbach}
\email{tim.berkelbach@gmail.com}
\affiliation
{Department of Chemistry, Columbia University, New York, New York 10027, USA}
\affiliation
{Center for Computational Quantum Physics, Flatiron Institute, New York, New York 10010, USA}
\author{Hong-Zhou Ye}
\email{hzyechem@gmail.com}
\affiliation
{Department of Chemistry, Columbia University, New York, New York 10027, USA}

\begin{abstract}
    We present an algorithm and implementation of integral-direct, density-fitted Hartree-Fock (HF) and second-order M{\o}ller-Plesset perturbation theory (MP2) for periodic systems.
    The new code eliminates the formerly prohibitive storage requirements and allows us to study systems one order of magnitude larger than before at the periodic MP2 level.
    We demonstrate the significance of the development by studying the benzene crystal in both the thermodynamic limit and the complete basis set limit, for which we predict an MP2 cohesive energy of $-72.8$~kJ/mol, which is about $10$--$15$~kJ/mol larger in magnitude than all previously reported MP2 calculations.
    Compared to the best theoretical estimate from literature, several modified MP2 models approach chemical accuracy in the predicted cohesive energy of the benzene crystal and hence may be promising cost-effective choices for future applications on molecular crystals.
\end{abstract}

\maketitle


\section{Introduction}

Recent years have witnessed a rapid growth of interest in leveraging systematically improvable wavefunction-based quantum chemistry methods to study challenging problems in materials science \cite{Marsman09JCP,Maschio11JPCA,Muller12PCCP,DelBen12JCTC,DelBen13JCTC,Booth13Nature,Yang14Science,McClain17JCTC,Schafer17JCP,Gruber18PRX,Zhang19FM,Wang20JCTC,Lau21JPCL,Lange21JCP,Wang21JCTC,Mihm21NCS,Wang21JACSAu,Nusspickel22PRX,Neufeld22arXiv}.
These simulations, often performed using periodic boundary conditions, are computationally expensive because of the large simulation cells or dense $k$-point meshes needed to reach the thermodynamic limit \cite{Gruneis10JCP,Gruber18PRX,Mihm21NCS,Neufeld22arXiv} (TDL) and the large one-particle basis sets needed to reach the complete basis set (CBS) limit \cite{Marsman09JCP,Gruneis11JCTC,Shepherd12PRB,Booth16JCP,Callahan21JCP,Lee21JCP,Ye22JCTC}.
As in molecular calculations, the evaluation and storage of the electron-repulsion integrals (ERIs) represent a major computational bottleneck \cite{Schwegler96JCP,Challacombe97JCP,Ochsenfeld98JCP,Shao00CPL} in Hartree-Fock \cite{Roothaan51RMP} (HF) and low-order perturbation (e.g.,~the second-order M{\o}ller-Plesset perturbation theory \cite{Moller34PR}, MP2) calculations,
including simulations using Kohn-Sham density functional theory \cite{Hohenberg64PR,Kohn65PR} (KS-DFT) with hybrid \cite{Becke93JCP,Adamo99JCP,Heyd03JCP} and double-hybrid \cite{Grimme06JCP,Zhang09PNAS,KozuchJPCC,Kozuch11PCCP} exchange-correlation functionals.
In Ref.~\citenum{Sun17JCP}, the commonly used density fitting (DF) technique \cite{Whitten73JCP,Dunlap79JCP,Mintmire82PRA} was adapted for periodic systems to reduce the computational cost of handling the periodic ERIs.
The resulting implementation in the PySCF software package~\cite{Sun18WIRCMS,Sun20JCP} has been used in many applications \cite{Wang20JCTC,Wang21JCTC,Zhu21JCTC,Zhu21PRX,Nusspickel22PRX}.

\begin{figure}[!b]
    \centering
    \includegraphics[width=0.9\linewidth]{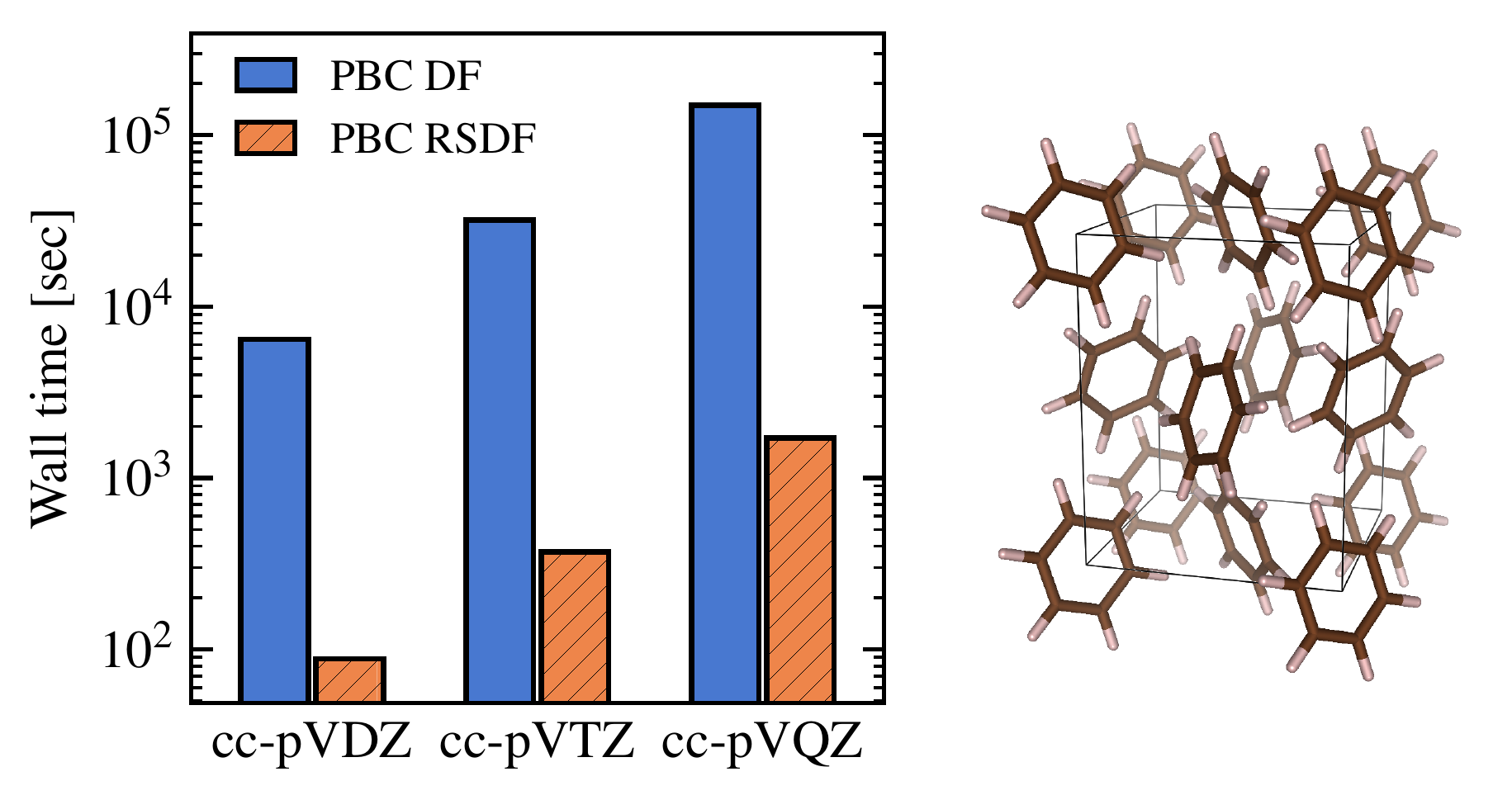}
    \caption{Wall time for calculating the DF integrals for the benzene crystal, whose unit cell is shown on the right, with $\Gamma$-point Brillouin zone sampling.
    The recently developed RSDF \cite{Ye21JCPa} (orange) algorithm accelerates the previous DF implementation (blue) by up to two orders of magnitude.
    All calculations are performed using PySCF on a single node with $16$ CPU cores.
    }
    \label{fig:df_timing_cmp}
\end{figure}

This previous implementation of periodic DF~\cite{Sun17JCP} is integral-indirect, meaning that the needed integrals are pre-computed and stored in memory or on disk for later use.
The resources needed to store the DF integrals grow quadratically with the number of $k$-points and cubically with the size of the unit cell or the basis set, preventing studies of large systems in the two limits.
An integral-\emph{direct} implementation that avoids storing all DF integrals at once is thus highly desirable but is hindered by the high computational cost of evaluating these integrals \cite{Sun17JCP}.
Recently, two of us introduced a range-separated DF \cite{Ye21JCPa} (RSDF) algorithm for fast evaluation of the DF integrals, which, when combined with efficient integral screening \cite{Ye21JCPb}, accelerates periodic DF by one to two orders of magnitude, as illustrated for the simulation of the benzene crystal in \cref{fig:df_timing_cmp}.

In this work, we leverage this significant speedup to enable an integral-direct implementation of periodic HF and MP2.
The development allows us to perform periodic HF and MP2 calculations for systems one order of magnitude larger than with the previous integral-indirect implementation.
We demonstrate the significance of this development by estimating the MP2 cohesive energy of the benzene crystal in both the TDL and the CBS limit.
A careful comparison to existing MP2 results in the literature \cite{Ringer08EJC,Maschio11JPCA,DelBen12JCTC} suggests that they may have large finite-size and/or basis set incompleteness errors, emphasizing the challenge and importance of reaching both the TDL and the CBS limit in correlated wavefunction-based simulations of materials.
We also show that various modified MP2 models\cite{Grimme03JCP,Jung04JCP,Distasio07MP,Tan17JCP} exhibit nearly chemical accuracy in the computed cohesive energy of the benzene crystal and hence may be promising for future applications on molecular crystals.

\section{Theory}

We start by briefly reviewing the formalism of periodic DF.
In periodic systems, the atom-centered Gaussian-type atomic orbitals (AOs) are translational symmetry-adapted
\begin{equation}
    \phi_{\mu}^{\bm{k}}(\bm{r})
        = \sum_{\bm{R}} \me^{\mi \bm{k}\cdot\bm{R}} \phi_{\mu}(\bm{r}-\bm{R})
\end{equation}
where the lattice summation runs over all unit cells in real space and $\bm{k}$ is one of the $N_k$ crystal momenta sampled from the first Brillouin zone \cite{McClain17JCTC}.
The periodic DF expands the AO product density in a second, auxiliary set of translational symmetry-adapted Gaussian basis functions $\chi_{P}^{\bm{k}}(\bm{r})$~\cite{Sun17JCP,Ye21JCPa}
\begin{equation}    \label{eq:df_expansion}
        \phi_{\mu}^{\bm{k}_1*}(\bm{r})\phi_{\nu}^{\bm{k}_2}(\bm{r})
        \approx \sum_{P}^{n_{\tr{aux}}} d_{P\mu\nu}^{\bm{k}_1\bm{k}_2}
        \chi_{P}^{\bm{k}_{12}}(\bm{r})
\end{equation}
so that the ERIs can be approximated as
\begin{equation}    \label{eq:eri_by_df_coeff}
    V_{\mu\nu\lambda\sigma}^{\bm{k}_1\bm{k}_2\bm{k}_3\bm{k}_4}
        \approx \sum_{P,Q}^{n_{\tr{aux}}} d_{P\mu\nu}^{\bm{k}_1\bm{k}_2}
        J_{PQ}^{\bm{k}_{34}} d_{Q\lambda\sigma}^{\bm{k}_3\bm{k}_4}
\end{equation}
where $J_{PQ}^{\bm{k}} = (\chi_{P}^{-\bm{k}} | \chi_{Q}^{\bm{k}})$ is a two-center Coulomb integral and the crystal momentum conservation requires that $\bm{k}_{12} \equiv -\bm{k}_1+\bm{k}_2 = -\bm{k}_{34} + \bm{G}$, where $\bm{G}$ is a reciprocal lattice vector.
The fitting coefficients are determined by solving a linear equation
\begin{equation}    \label{eq:df_eqn}
    \sum_{Q}^{n_{\tr{aux}}} J_{PQ}^{\bm{k}_{12}} d_{Q\mu\nu}^{\bm{k}_1\bm{k}_2}
        = V_{P\mu\nu}^{\bm{k}_1\bm{k}_2}
\end{equation}
which allows one to rewrite \cref{eq:eri_by_df_coeff} as
\begin{equation}
    V_{\mu\nu\lambda\sigma}^{\bm{k}_1\bm{k}_2\bm{k}_3\bm{k}_4}
        \approx \sum_{P,Q}^{n_{\tr{aux}}} V_{P\mu\nu}^{\bm{k}_1\bm{k}_2}
        [(\mat{J}^{\bm{k}_{34}})^{-1}]_{PQ} V_{Q\lambda\sigma}^{\bm{k}_3\bm{k}_4}
        = \sum_{P} \tilde{V}_{P\mu\nu}^{\bm{k}_1\bm{k}_2}
        \tilde{V}_{P\lambda\sigma}^{\bm{k}_3\bm{k}_4}
\end{equation}
where $V_{P\mu\nu}^{\bm{k}_1\bm{k}_2} = (\chi_{P}^{-\bm{k}_{12}}| \phi_{\mu}^{\bm{k}_1*}\phi_{\nu}^{\bm{k}_2})$,
$\tilde{\mat{V}}^{\bm{k}_1\bm{k}_2} = \mat{L}^{\bm{k}_{12}\dagger}\mat{V}^{\bm{k}_1\bm{k}_2}$,
and $\mat{L}^{\bm{k}}$ is the lower-triangular matrix from the Cholesky decomposition of $(\mat{J}^{\bm{k}})^{-1}$, i.e.,~$(\mat{J}^{\bm{k}})^{-1} = \mat{L}^{\bm{k}} \mat{L}^{\bm{k}\dagger}$.
To summarize, periodic DF factorizes the periodic four-center ERIs into periodic two-center and three-center Coulomb integrals, and this compression is responsible for the reduced storage requirements.

However, even with DF, storage is still the main computational bottleneck for large systems: storing the three-center Coulomb integrals $V_{P\mu\nu}^{\bm{k}_1\bm{k}_2}$ requires $O(N_k^2 n_{\tr{aux}} n_{\tr{AO}}^2)$ memory or disk space, i.e., it scales quadratically with the number of $k$-points and cubically with the size of the unit cell or the basis set.
The basic idea of an integral-direct implementation is to calculate the three-center integrals on-the-fly to avoid the high cost of storing them all at once.
\rvv{In this work, using periodic integral evaluation with RSDF, we calculate the integrals in blocks and batch one of the two AO indices, which we denote by $V_{P[\mu]\nu}^{\bm{k}_1\bm{k}_2}$.
The alternative choices to batch over the auxiliary function index or the $k$-points are considered in the Supporting Information, where we argue that batching over an AO index (as we do here) is best for calculations with large unit cells and small $k$-point meshes, but batching over $k$-points will be best for calculations with small unit cells and large $k$-point meshes (larger than $N_k \approx 5^3$ with a high-quality basis set).}

We first discuss our integral-direct implementation of periodic HF, which resembles the algorithms previously developed for molecular HF calculations \cite{Neese02CPL,Weigend02PCCP} but is made compatible here with the $k$-point symmetry that is unique to periodic systems.
\rvv{(See also Refs.~\citenum{Sun21arXiv,Sharma22arXiv,Lee22arXiv} for recent related developments in periodic exchange evaluation.)}
Our goal is to calculate the Coulomb and the exchange matrices, referred to as the $J$-build and $K$-build, in an integral-direct manner.
The discussion below assumes a spin-restricted mean-field state with crystalline orbitals (COs)
\begin{equation}
    \psi_{p}^{\bm{k}}(\bm{r})
        = \sum_{\mu}^{n_{\tr{AO}}} C_{\mu p}^{\bm{k}} \phi_{\mu}^{\bm{k}}(\bm{r})
\end{equation}
and the corresponding CO energies $\varepsilon_p^{\bm{k}}$.
(The common notation of $i,j,\cdots$ labelling $n_{\tr{occ}}$ occupied COs, $a,b,\cdots$ labelling $n_{\tr{vir}}$ virtual COs, and $p,q,\cdots$ labelling unspecified COs, will be used throughout the paper.)
The extension to a general state that breaks spin symmetry is straightforward.

With DF, the Coulomb matrix is calculated as
\begin{equation}    \label{eq:Jbuild_pass2}
    J_{\mu\nu}^{\bm{k}}
        = \sum_{P}^{n_{\tr{aux}}} V_{P\mu\nu}^{\bm{k}\bm{k}} \tilde{v}_{P}
\end{equation}
where $\tilde{\bm{v}} = (\mat{J}^{\bm{0}})^{-1}\bm{v}$,
\begin{equation}    \label{eq:Jbuild_pass1}
    v_{P}
        = \frac{1}{N_k} \sum_{\bm{k}}^{N_k} \sum_{\lambda\sigma}^{n_{\tr{AO}}} V_{P\lambda\sigma}^{\bm{k}\bm{k}} D_{\sigma\lambda}^{\bm{k}},
\end{equation}
and $D_{\sigma\lambda}^{\bm{k}}$ is the HF density matrix.
The intermediates $\bm{v}$ and $\tilde{\bm{v}}$ are of size $O(n_{\tr{aux}})$ and can always be held in memory.
We note that only three-center integrals that are diagonal in $\bm{k}$ are needed,
but the cubic scaling with the unit cell size or the basis set size is unchanged and can still be the bottleneck for large unit cells and/or large basis sets.
To that end, we perform the tensor contractions in \cref{eq:Jbuild_pass2,eq:Jbuild_pass1} in blocks by batching one of the AO indices,
\begin{equation}    \label{eq:Jbuild_pass2_batch}
    J_{[\mu]\nu}^{\bm{k}}
        = \sum_{P}^{n_{\tr{aux}}} V_{P[\mu]\nu}^{\bm{k}\bm{k}} \tilde{v}_{P}
\end{equation}
for \cref{eq:Jbuild_pass2} and
\begin{equation}    \label{eq:Jbuild_pass1_batch}
    v_{P}
        = \sum_{[\lambda]} \bigg( \frac{1}{N_k} \sum_{\bm{k}}^{N_k}
        \sum_{\lambda\in[\lambda]} \sum_{\sigma}^{n_{\tr{AO}}}
        V_{P\lambda\sigma}^{\bm{k}\bm{k}} D_{\sigma\lambda}^{\bm{k}} \bigg)
\end{equation}
for \cref{eq:Jbuild_pass1}.
The batching here introduces no extra computational cost but simply avoids storing the full $V_{P\mu\nu}^{\bm{k}\bm{k}}$ tensors.

For the exchange matrix, we adapt the occupied orbital-based $K$-build algorithm \cite{Weigend02PCCP,Koppl16JCTC} for periodic calculations with DF,
\begin{equation}    \label{eq:Kbuild_WW}
    K_{\mu\nu}^{\bm{k}_1}
        = \frac{1}{N_k} \sum_{\bm{k}_2}^{N_k}
        \sum_{i}^{n_{\tr{occ}}} \sum_{P}^{n_{\tr{aux}}}
        \tilde{W}_{P\mu i}^{\bm{k}_1\bm{k}_2}
        \tilde{W}_{P\nu i}^{\bm{k}_1\bm{k}_2*}
\end{equation}
where $\tilde{\mat{W}}^{\bm{k}_1\bm{k}_2} = \mat{L}^{\bm{k}_{12}\dagger} \mat{W}^{\bm{k}_1\bm{k}_2}$ and
\begin{equation}    \label{eq:WPmui_batch}
    W_{P[\mu] i}^{\bm{k}_1\bm{k}_2}
        = \sum_{\sigma}^{n_{\tr{AO}}} V_{P[\mu]\sigma}^{\bm{k}_1\bm{k}_2}
        C_{\sigma i}^{\bm{k}_2} \sqrt{n_{i}^{\bm{k}_2}}
\end{equation}
with $n_{i}^{\bm{k}}$ the CO occupation number (i.e.,~$2$ for a spin-restricted state).
Like for the $J$-build, we avoid the storage of the entire $V_{P\mu\sigma}^{\bm{k}_1\bm{k}_2}$ tensor in the half-transformation (\ref{eq:WPmui_batch}) by batching over an AO index.
The alternative that batches the $\sigma$ index in \cref{eq:WPmui_batch} is suboptimal because it requires repeated tensor addition to accumulate the results.
Because $W_{P\mu i}^{\bm{k}_1\bm{k}_2}$ is smaller than $V_{P\mu\nu}^{\bm{k}_1\bm{k}_2}$ by a factor of $n_{\tr{AO}} / n_{\tr{occ}}$, it can be stored in its entirety for significantly larger systems.
When it can be stored in memory, this completes our description of a fully direct periodic $K$-build.

When $W_{P\mu i}^{\bm{k}_1\bm{k}_2}$ does not fit in memory but does fit on disk, we use a semi-direct algorithm. In this case, we store $W_{P\mu i}^{\bm{k}_1\bm{k}_2}$ on disk and loaded into memory in blocks by batching the $i$ index in \cref{eq:Kbuild_WW},
\begin{equation}    \label{eq:Kbuild_WW_batch}
    K_{\mu\nu}^{\bm{k}_1}
        = \sum_{[i]} \bigg( \frac{1}{N_k} \sum_{\bm{k}_2}^{N_k}
        \sum_{i \in [i]} \sum_{P}^{n_{\tr{aux}}}
        \tilde{W}_{P\mu i}^{\bm{k}_1\bm{k}_2}
        \tilde{W}_{P\nu i}^{\bm{k}_1\bm{k}_2*} \bigg).
\end{equation}
If necessary, \cref{eq:Kbuild_WW_batch} can be used in a fully direct manner (i.e., using only memory), but this increases the computational cost compared to the semi-direct algorithm because each batch of the half-transformed integrals, $W^{\bm{k}_1\bm{k}_2}_{P\mu[i]}$, requires evaluating the entire set of three-center integrals $V_{P\mu\nu}^{\bm{k}_1\bm{k}_2}$.
Whether the semi-direct approach is more efficient than the fully direct alternative depends on the relative cost of integral evaluation compared to writing to and reading from disk.
For the current RSDF implementation, we found by numerical tests that the integral evaluation is still the computational bottleneck,
and thus we use the semi-direct approach throughout this work for the $K$-build.
We note that the situation may change depending on the compute architecture, available resources, or with further development of periodic integral evaluation (see e.g.,~ref \citenum{Sharma21JCTC}).

Lastly, we discuss the integral-direct implementation of periodic MP2.
The correlation energy for periodic MP2 is
\begin{equation}    \label{eq:EMP2}
    E^{\tr{MP2,c}}
        = -\frac{1}{N_k^3} \sum_{\bm{k}_1\bm{k}_2\bm{k}_3}^{N_k}
        \sum_{abij}
        \frac{V_{aibj}^{\bm{k}_1\bm{k}_2\bm{k}_3\bm{k}_4*}
            (2 V_{aibj}^{\bm{k}_1\bm{k}_2\bm{k}_3\bm{k}_4} -
            V_{biaj}^{\bm{k}_3\bm{k}_2\bm{k}_1\bm{k}_4})}
        {\varepsilon_a^{\bm{k}_1} - \varepsilon_i^{\bm{k}_2} +
        \varepsilon_b^{\bm{k}_3} - \varepsilon_j^{\bm{k}_4}}
\end{equation}
where $\bm{k}_4 = \bm{k}_1 - \bm{k}_2 + \bm{k}_3 + \bm{G}$ by crystal momentum conservation.
With DF, the transformed ERIs are approximated by three-index tensors
\begin{equation}
    V_{aibj}^{\bm{k}_1\bm{k}_2\bm{k}_3\bm{k}_4}
        \approx \sum_{P}^{n_{\tr{aux}}} \tilde{U}^{\bm{k}_1\bm{k}_2}_{Pai}
        \tilde{U}^{\bm{k}_3\bm{k}_4}_{Pbj}
\end{equation}
where $\tilde{\mat{U}}^{\bm{k}_1\bm{k}_2} = \mat{L}^{\bm{k}_{12}\dagger} \mat{U}^{\bm{k}_1\bm{k}_2}$ and
\begin{equation}    \label{eq:UPai}
    U^{\bm{k}_1\bm{k}_2}_{Pai}
        = \sum_{\mu}^{n_{\tr{AO}}}
        \bigg(
            \sum_{\nu}^{n_{\tr{AO}}} V_{P\mu\nu}^{\bm{k}_1\bm{k}_2}
            C_{\nu i}^{\bm{k}_2}
        \bigg)
        C_{\mu a}^{\bm{k}_1*}
        = \sum_{\mu}^{n_{\tr{AO}}} W_{P\mu i}^{\bm{k}_1\bm{k}_2} C_{\mu a}^{\bm{k}_1*}
\end{equation}
are transformed three-center integrals, where we used \cref{eq:WPmui_batch} ($n^{\bm{k}}_i = 1$ here) for the second equality.
The half-transformed integrals $W_{P\mu i}^{\bm{k}_1\bm{k}_2}$ in \cref{eq:UPai} can be computed as discussed above for the $K$-build and stored on disk.
These integrals are then loaded into memory in blocks by batching the $i$ index for the second transform in \cref{eq:UPai};
the alternative that batches the $\mu$ index is suboptimal due to the repeated tensor addition for accumulating the results.
The $U_{Pai}^{\bm{k}_1\bm{k}_2}$ tensors are marginally smaller than $W_{P\mu i}^{\bm{k}_1\bm{k}_2}$ (by a factor of $n_{\tr{AO}}/n_{\tr{vir}}$), and therefore have similar storage requirements.
If $U_{Pai}^{\bm{k}_1\bm{k}_2}$ exceeds the available disk space, we compute it in blocks by batching the $i$ index, $U^{\bm{k}_1\bm{k}_2}_{Pa[i]}$, and compute the MP2 energy in blocks accordingly
\begin{equation}    \label{eq:EMP2_batch}
    E^{\tr{MP2,c}}
        = \sum_{[i]} \sum_{[j]} \bigg(
            - \frac{1}{N_k^3} \sum_{\bm{k}_1\bm{k}_2\bm{k}_3}^{N_k}
            \sum_{i\in[i]} \sum_{j\in[j]} \sum_{ab}^{n_{\tr{vir}}}
            \cdots
        \bigg)
\end{equation}
where the summand is the same as that in \cref{eq:EMP2} and omitted here.
Although not explored in this work, the MP2 one-particle reduced density matrix, which is useful in various reduced-scaling correlated methods based on MP2 natural orbitals \cite{Jensen88JCP,Landau10JCP,Gruneis11JCTC,Kumar17JCPA,Guo18JCP,Nagy19JCTC,Lange20MP}, can be evaluated in essentially the same manner.
\rvv{Additional approximations such as the Laplace transform that have been shown to further reduce the computational cost of canonical periodic MP2 calculations \cite{Schafer17JCP} will be explored in future work.}

\section{Computational details}

The integral-direct algorithms presented above for periodic HF and MP2 calculations with DF are implemented in the PySCF software package \cite{Sun18WIRCMS,Sun20JCP} which uses libcint \cite{Sun15JCC} for calculating atomic integrals.
We demonstrate the impact of our integral-direct algorithms by estimating the MP2 cohesive energy of the benzene crystal in both the TDL and the CBS limit.
The cohesive energy of the benzene crystal has been well-studied in the literature using molecular codes via the truncated many-body expansion (MBE) \cite{Stoll92PRB,Hirata05MP,Paulus06PR,Kamiya08JCP} with several correlated wavefunction methods \cite{Schweizer06JCTC,Yang14Science} including MP2 \cite{Ringer08EJC}.
Two different periodic MP2 calculations have also been reported \cite{Maschio11JPCA,DelBen12JCTC}, showing good agreement with each other but differing from MBE results \cite{Ringer08EJC} by about $7$~kJ/mol.
Here, we leverage the power of our integral-direct implementations to investigate these discrepancies through our own careful investigation of finite-size and basis set errors, ultimately finding an MP2 cohesive energy that is larger in magnitude than any of these previous studies.

All calculations reported below were performed using PySCF on a single compute node with $384$ GB of memory and $1$ TB of disk space.
The Brillouin zone is sampled by uniform $k$-point meshes including the $\Gamma$-point.
Finite-size errors associated with the divergence of the HF exchange integral at $G=0$ are handled using a Madelung constant correction~\cite{Paier06JCP,Broqvist09PRB,Sundararaman13PRB}.
With this treatment, both the HF energy and the MP2 correlation energy exhibit a $1/N_k$ asymptotic convergence to the TDL (i.e.,~$N_k = \infty$) and can hence be extrapolated using the following two-point formula
\begin{equation} \label{eq:tdl_extrap}
    E(\infty)
        = \frac{ N_{k,2}^{-1} E(N_{k,1}) - N_{k,1}^{-1} E(N_{k,2}) }
        {N_{k,2}^{-1} - N_{k,1}^{-1}}
\end{equation}
for sufficiently large $N_{k,1}$ and $N_{k,2}$.
We denote an extrapolation based on \cref{eq:tdl_extrap} $(N_{k,1},N_{k,2})$.

\section{Results and discussion}

\begin{figure}[!t]
    \centering
    \includegraphics[width=1.0\linewidth]{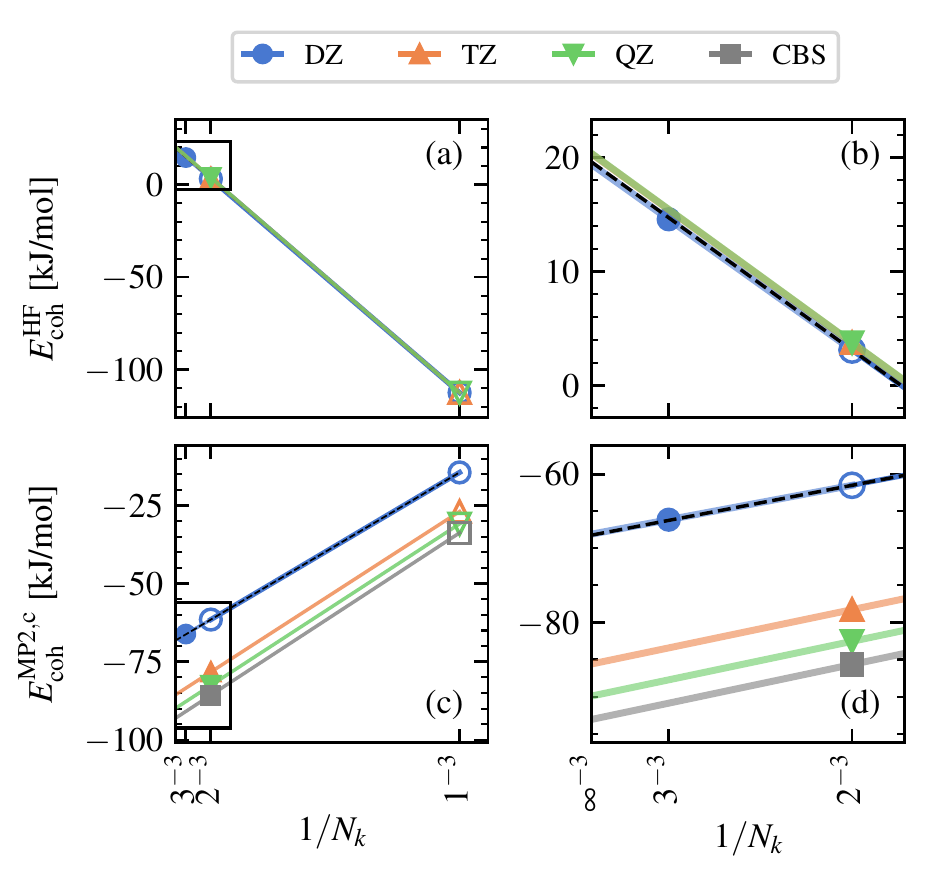}
    \caption{
    Thermodynamic limit convergence of the HF cohesive energy (a) and the MP2 correlation energy contribution to the cohesive energy (c) of the benzene crystal (CSD code BENZEN01) using different basis sets.
    Panels (b) and (d) provide zoom-in views of the corresponding region in (a) and (c) indicated by black rectangles.
    Hollow symbols correspond to calculations that can be performed using the previous integral-indirect implementation, while filled symbols are calculations made possible by the integral-direct implementation developed in this work.
    For each basis set, the TDL extrapolation based on the two largest calculations [$(2^3,3^3)$ for DZ and $(1^3,2^3)$ for others] is shown as a solid line of the corresponding color.
    For DZ, the $(1^3,2^3)$ TDL extrapolation is also shown as a black dashed line.
    }
    \label{fig:BENZEN01_hf_mp2_ecoh}
\end{figure}

We first calculate the cohesive energy of the benzene crystal for the $138$ K lattice geometry \cite{Bacon64PRSA} [code BENZEN01 in the Cambridge Structure Database \cite{Groom16ACSB} (CSD)] using the all-electron cc-pV$X$Z (henceforth referred to as $X$Z) basis sets \cite{Dunning89JCP} up to QZ.
As shown in \cref{fig:df_timing_cmp}, each unit cell contains four benzene molecules, 168 electrons, and 456, 1056, and 2040 AOs with the DZ, TZ, and QZ basis sets, respectively.
The corresponding cc-pV$X$Z-JKFIT basis sets \cite{Weigend02PCCP} are used for DF.
The $1s$ core electrons of carbon are kept frozen in the MP2 calculations.
The same lattice geometry and similar basis sets were used in previous MBE calculations \cite{Schweizer06JCTC,Ringer08EJC,Yang14Science}.
The $k$-point convergence of the cohesive energy from our periodic HF and MP2 calculations is shown in \cref{fig:BENZEN01_hf_mp2_ecoh} for different basis sets.
For MP2, an estimate of the CBS limit of a given $k$-point mesh is obtained by a $1/X^3$ extrapolation using the TZ ($X=3$) and the QZ ($X=4$) results of the same $k$-point mesh.
For HF, the change of the cohesive energy from TZ to QZ is less than $0.03$ kJ/mol for all $k$-point meshes.
Thus, the QZ HF results are taken as the CBS limit without further extrapolation.

With the previous integral-indirect code, we can compute the cohesive energies using $N_k = 1^3$ and $2^3$ with the DZ basis set, but only using $N_k = 1^3$ with the TZ and QZ basis sets, all of which are marked by hollow symbols in \cref{fig:BENZEN01_hf_mp2_ecoh}.
Therefore, the TDL extrapolation using \cref{eq:tdl_extrap} can only be performed with DZ (black dashed lines) and gives a cohesive energy of $19.6$ kJ/mol for HF and $-48.6$ kJ/mol for MP2, respectively.
The quality of this $(1^3,2^3)$ TDL extrapolation is, however, questionable due to the use of relatively small $k$-point meshes.
In addition, the MP2 cohesive energy at $\Gamma$-point obtained using the DZ basis set is about $20$ kJ/mol higher than the estimated CBS limit as shown in \cref{fig:BENZEN01_hf_mp2_ecoh}(c), indicating a large basis set incompleteness error.
A simple composite estimate, based on these minimal data points, suggests an MP2 cohesive energy of $-67.9$ kJ/mol in the combined TDL and CBS limit, which underestimates our best estimate by about $5$ kJ/mol (\textit{vide infra}).

The integral-direct code developed in this work allows us to obtain the cohesive energies for $k$-point meshes one order of magnitude larger than before, i.e.,~$N_k = 3^3$ with DZ and $N_k = 2^3$ with TZ and QZ, as marked by filled symbols in \cref{fig:BENZEN01_hf_mp2_ecoh}.
For DZ, a $(2^3,3^3)$ TDL extrapolation using \cref{eq:tdl_extrap} (blue solid lines) gives a cohesive energy of $19.4$ kJ/mol for HF and $-48.7$ kJ/mol for MP2, respectively, which agree very well with the $(1^3,2^3)$ TDL extrapolation discussed above [see also the overlay of the blue solid line and the black dashed line in \cref{fig:BENZEN01_hf_mp2_ecoh}(b,d)].
The nearly quantitative agreement justifies a $(1^3,2^3)$ TDL extrapolation for larger basis sets followed by a composite correction from the difference between the $(2^3,3^3)$ and $(1^3,2^3)$ TDL extrapolations of DZ (which we denote by $\Delta \tr{DZ}$).
The obtained cohesive energies in the TDL for various basis sets and the estimated CBS limit are listed in \cref{tab:BENZEN01_hf_mp2_ecoh}, along with results from the literature for comparison.

\begin{table}[!t]
    \centering
    \caption{Cohesive energy of the benzene crystal.
    Results are reported for the $138$ K lattice structure\cite{Bacon64PRSA} (CSD code BENZEN01) unless otherwise specified.}
    \label{tab:BENZEN01_hf_mp2_ecoh}
    \begin{threeparttable}
        \begin{tabular*}{\linewidth}{l @{\extracolsep{\fill}} llccl}
            \hline\hline
            & \multirow{2}*{Basis set} & \multirow{2}*{TDL} &
                \multicolumn{2}{c}{$E_{\tr{coh}}$ [kJ/mol]} & \\
            \cmidrule(lr){4-5}
            & & & HF & MP2 & \\
            \hline

            MBE & & & & & \\

            & cc-pV5Z & & $20.2$ & N/A & ref \citenum{Yang14Science} \\
            & (T,Q)-CBS & & N/A & $-64.0$ & ref \citenum{Ringer08EJC} \\
            & & & & & \\

            Periodic & & & & \\

            & cc-pVDZ & $(2^3,3^3)$ & $19.4$ & $-48.7$ & this work \\
            & cc-pVTZ & $(1^3,2^3) + \Delta\tr{DZ}$ & $20.1$ & $-65.5$ & this work \\
            & cc-pVQZ & $(1^3,2^3) + \Delta\tr{DZ}$ & $20.2$ & $-69.8$ & this work \\
            & (T,Q)-CBS & $(1^3,2^3) + \Delta\tr{DZ}$ & $20.2$ & $-72.8$ & this work \\

            & & & & & \\

            & p-aug-6-31G**\tnote{a} & N/A & N/A & $-56.6$\tnote{b} & ref \citenum{Maschio11JPCA}\tnote{c} \\
            & p-aug-6-31G**\tnote{a} & $(1^3,2^3)$ & $20.0$ & $-67.9$ & this work\tnote{c} \\

            & & & & & \\

            & cc-TZVP\tnote{d} & $2\times1\times2$\tnote{e} & $21.2$ & $-58.7$ & ref \citenum{DelBen12JCTC}\tnote{c} \\
            & cc-TZVP\tnote{d} & $(1^3,2^3)$ & $20.5$ & $-69.5$ & this work\tnote{c} \\
            \hline
        \end{tabular*}
        \begin{tablenotes}
            \footnotesize
            \item[a] The diffuse $p$ function for H and $d$ function for C from the aug-cc-pVDZ basis set are added 6-31G**.
            \item[b] Using local MP2 \cite{Pisani05JCP} (LMP2).
            \item[c] Using the 123 K lattice structure \cite{Jeffrey87PRSA} (CSD code BENZEN07).
            \item[d] Using the GTH pseudopotential optimized for HF \cite{HutterPP}.
            \item[e] Using the truncated Coulomb potential \cite{Spencer08PRB} for HF.
        \end{tablenotes}
    \end{threeparttable}
\end{table}

The cohesive energy from our periodic HF calculations in the CBS limit ($20.2$ kJ/mol) agrees quantitatively with that obtained from a MBE truncated to tetramers~\cite{Yang14Science}.
Our estimated MP2 cohesive energy in the CBS limit ($-72.8$ kJ/mol) is about $9$ kJ/mol larger in magnitude than the MBE result in ref \citenum{Ringer08EJC}, which considered only dimer interactions.
We attribute the difference to the neglect of contributions from trimers and tetramers, which have been shown to cause a sizable error for the benzene crystal \cite{Yang14Science}.

Also listed in \cref{tab:BENZEN01_hf_mp2_ecoh} are the cohesive energies from two periodic MP2 studies in literature \cite{Maschio11JPCA,DelBen12JCTC} for the $123$ K lattice structure \cite{Jeffrey87PRSA} (CSD code BENZEN07).
Ref \citenum{Maschio11JPCA} uses a partially augmented 6-31G** (p-aug-6-31G**) basis set and obtains an MP2 cohesive energy of $-56.6$ kJ/mol, while ref \citenum{DelBen12JCTC} uses a TZ-quality basis set (cc-TZVP) and Goedecker-Teter-Hutter (GTH) pseudopotentials \cite{Goedecker96PRB,Hartwigsen98PRB} and predicts a similar value of $-58.7$ kJ/mol.
Despite the reasonable agreement between them, these values are noticeably smaller in magnitude, by up to $16$ kJ/mol, than our best estimate in the TDL and the CBS limit.
We repeated our MP2 calculations using the same basis sets and lattice structure as in these previous works, but extrapolated to the TDL based on the $(1^3,2^3)$ scheme established above.
As shown in \cref{tab:BENZEN01_hf_mp2_ecoh}, the difference between our MP2 cohesive energies and the literature values suggests that the latter have a finite-size error of about $11$ kJ/mol.
The difference from our best estimate in the CBS limit reveals a basis set incompleteness error of about $3$ and $5$ kJ/mol for the cc-TZVP and the p-aug-6-31G** basis sets, respectively (we have numerically confirmed that the two crystal structures have cohesive energies that differ by less than $1$~kJ/mol).
These comparisons demonstrate the challenge of reaching the combined TDL and CBS limit and the value of our integral-direct algorithms that enable calculations with large $k$-point meshes and large basis sets.

\begin{figure}[!t]
    \centering
    \includegraphics[width=0.7\linewidth]{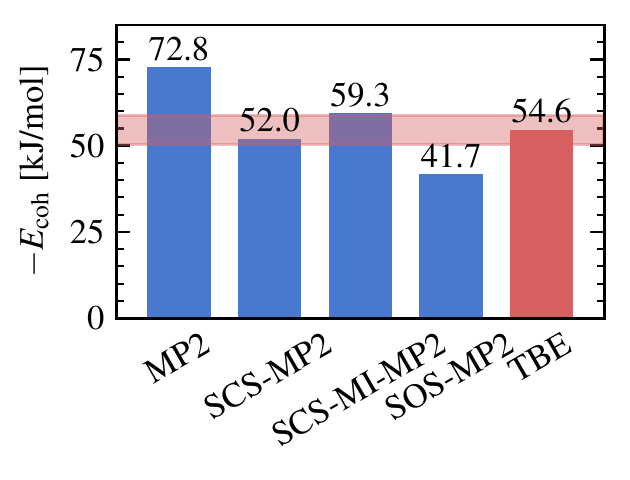}
    \caption{Cohesive energy of the benzene crystal (138 K lattice structure) computed from MP2 and its empirical modifications in both the TDL [$(1^3,2^3) + \Delta \tr{DZ}$] and the CBS limit [(T,Q)].
    The theoretical best estimate (TBE) from ref \citenum{Yang14Science} is plotted for comparison.
    The red shaded area indicates $\pm 1$ kcal/mol from the TBE.}
    \label{fig:ecoh_scs_mp2}
\end{figure}

Finally, we gauge the performance of various empirically modified MP2 models that are commonly used for molecules \cite{Grimme03JCP,Jung04JCP,Distasio07MP,Tan17JCP}, \rvv{and which we have found, in forthcoming work from our group~\cite{Goldzak22UP}, to significantly improve the cohesive properties of covalently bound semiconductors and insulators.}
These models are based on scaling the correlation energy of different spin components [i.e.,~same-spin (SS) and opposite-spin (OS)] with different coefficients
\begin{equation}    \label{eq:SCSMP2}
    E^{\tr{MP2,c}}_{\tr{modified}}(c_{\tr{SS}},c_{\tr{OS}})
        = c_{\tr{SS}} E^{\tr{MP2,c}}_{\tr{SS}} + c_{\tr{OS}} E^{\tr{MP2,c}}_{\tr{OS}}
\end{equation}
where the unmodified MP2 model is recovered for $c_{\tr{SS}} = c_{\tr{OS}} = 1$.
\Cref{fig:ecoh_scs_mp2} shows the benzene crystal cohesive energy computed from three such models in the TDL and the CBS limit, along with the theoretical best estimate (TBE) from ref \citenum{Yang14Science} for comparison.
All modified MP2 models correct for the known overestimation of the dispersion interaction by unmodified MP2 \cite{Cybulski07JCP,Hebelmann08JCP}.
The general-purpose spin-component-scaled (SCS) model \cite{Grimme03JCP} and the SCS-molecular interaction (SCS-MI) model \cite{Distasio07MP} parameterized for reproducing the CCSD(T) \cite{Raghavachari89CPL} intermolecular interactions both give results within chemical accuracy ($1$ kcal/mol or $4.2$ kJ/mol; see the red shaded area in \cref{fig:ecoh_scs_mp2}),
while the scaled-opposite-spin (SOS) model \cite{Jung04JCP} significantly underestimates the TBE by about $13$ kJ/mol, which is consistent with previous literature results \cite{Lochan05JPCA,Distasio07MP}.

\section{Conclusion}

To conclude, in this work we reported an integral-direct implementation of periodic HF and MP2 with DF, which is made possible by our recent developments in periodic DF integral evaluation \cite{Ye21JCPa,Ye21JCPb}.
The development enables us to study systems one order of magnitude larger than before and allowed us to estimate the MP2 cohesive energy of the benzene crystal in both the TDL and the CBS limit, which in turn corrects the previously reported MP2 results from the literature.
Several modified MP2 models were shown to exhibit nearly chemical accuracy for the benzene crystal cohesive energy, which suggests that modified MP2 models and the closely related double-hybrid KS-DFT \cite{Grimme06JCP,Zhang09PNAS,KozuchJPCC,Kozuch11PCCP,Stein20Mol,Wang21JACSAu} may be cost-effective choices for crystal structure prediction.

The integral-direct code developed in this work has essentially eliminated the storage bottlenecks of large, periodic electronic structure calculations at the presented levels of theory.
However, it does not lower their computational scaling, which is now the bottleneck that precludes larger calculations.
For truly large-scale applications, local approximations in one form or another \cite{Maschio07PRB,Usvyat07PRB,Pinski15JCP,Koppl16JCTC,Tew18JCP,Wang20JCP} are necessary, and we expect that the work presented here will be essential in the benchmarking and development of those methods.

\section*{Acknowledgements}

We thank Dr.~Xiao Wang for helpful discussions.
This work was supported by the National Science Foundation under
Grant No.~DGE-1644869 (S.J.B.) and Grant No.~OAC-1931321 (H.-Z.Y.).
We acknowledge computing resources from Columbia University's
Shared Research Computing Facility project, which is supported by NIH Research
Facility Improvement Grant 1G20RR030893-01, and associated funds from the New
York State Empire State Development, Division of Science Technology and
Innovation (NYSTAR) Contract C090171, both awarded April 15, 2010. The Flatiron
Institute is a division of the Simons Foundation.

\section*{Supporting Information}

See the supporting information for (i) a review of periodic DF and the RSDF algorithm and (ii) a comparison of different strategies for batching the three-center integrals.

\section*{Data availability statement}
The data that support the findings of this study are available from the
corresponding author upon reasonable request.

\bibliography{refs}

\raggedbottom

\end{document}


\maketitle

\tableofcontents

\vspace{3em}

Note: figures and equations appearing in the main text will be referred as
``Fig.\ Mxxx'' and ``Eq.\ Mxxx'' in this Supplementary Material document.

\clearpage

\section{Range-separated density fitting (RSDF)}

We begin with a brief review of periodic Gaussian density fitting~\cite{Sun17JCP} and its efficient implementation with
range separation (RSDF) \cite{Ye21JCPa}, whose special structure puts some constraints on the way one can batch the evaluation of the DF 3c integrals (see \cref{sec:cmp_batch}).
In RSDF, we break the DF 3c integrals into a short-range (SR) part and a long-range (LR) part
\begin{equation}    \label{eq:3c_RSDF}
    \mat{V}^{\bm{k}_1\bm{k}_2}
        = \mat{V}^{\tr{SR},\bm{k}_1\bm{k}_2} + \mat{V}^{\tr{LR},\bm{k}_1\bm{k}_2}
\end{equation}
which correspond to the integrals of the SR part and the LR part of the Coulomb potential
\begin{equation}
    \frac{1}{r_{12}}
        = \underbrace{
            \frac{\tr{erfc}(\omega r_{12})}{r_{12}}
        }_{v^{\tr{SR}}(r_{12};\omega)} +
        \underbrace{
            \frac{\tr{erf}(\omega r_{12})}{r_{12}}
        }_{v^{\tr{LR}}(r_{12};\omega)}.
\end{equation}

The SR 3c integrals are evaluated in real-space by a lattice summation
\begin{equation}    \label{eq:SR3c}
    V^{\tr{SR},\bm{k}_1\bm{k}_2}_{P\mu\nu}(\omega)
        = \sum_{\bm{m}_1\bm{m}_2}^{N_{\tr{cell}}}
        \me^{-\mi\bm{k}_1\cdot\bm{m}_1} \me^{\mi\bm{k}_2\cdot\bm{m}_2}
        V_{P\mu\nu}^{\bm{0}\bm{m}_1\bm{m}_2}(\omega)
\end{equation}
where
\begin{equation}    \label{eq:MolSR3c}
    V_{P\mu\nu}^{\bm{0}\bm{m}_1\bm{m}_2}(\omega)
        = \iint\md\bm{r}_1\md\bm{r}_2\, \chi_{P}(\bm{r}_1) v^{\tr{SR}}(r_{12};\omega)
        \phi_{\mu}(\bm{r}_2-\bm{m}_1) \phi_{\nu}(\bm{r}_2-\bm{m}_2).
\end{equation}
is the regular molecular SR 3c integral with the two AOs shifted to cells $\bm{m}_1$ and $\bm{m}_2$, respectively.
Therefore, the cost of evaluating the SR 3c integrals has two parts: $O(N_{\tr{cell}} n_{\tr{aux}} n_{\tr{AO}}^2)$ for evaluating the molecular integrals (\ref{eq:MolSR3c}) and $O(N_k^3 n_{\tr{aux}} n_{\tr{AO}}^2)$ for the contraction in \cref{eq:SR3c}.

The LR 3c integrals are evaluated in reciprocal space using a plane wave (PW) basis (primed summation indicates that $\bm{G}+\bm{k}_{12} \neq \bm{0}$)
\begin{equation}    \label{eq:LR3c}
    V^{\tr{LR},\bm{k}_1\bm{k}_2}_{P\mu\nu}(\omega)
        = 4\pi \sum_{\bm{G}}^{N_{\tr{PW}}}{}'
        \frac{
            \me^{-|\bm{G}+\bm{k}_{12}|^2/(4\omega^2)}
        }{|\bm{G}+\bm{k}_{12}|^2}
        \tilde{\chi}^{\bm{k}_{12}}_{P}(-\bm{G})
        \tilde{\rho}_{\mu\nu}^{\bm{k}_1\bm{k}_2}(\bm{G})
\end{equation}
where
\begin{equation}    \label{eq:AFT_aux}
    \tilde{\chi}_{P}^{\bm{k}_{12}}(\bm{G})
        = \int\md\bm{r}\, \chi_{P}^{\bm{k}_{12}}(\bm{r})
        \me^{-\mi(\bm{k}_{12}+\bm{G})\cdot\bm{r}}
\end{equation}
is the analytical Fourier transform (AFT) of the auxiliary basis function, whose evaluation cost is negligible, and
\begin{equation}    \label{eq:AFT_AOpair}
    \tilde{\rho}_{\mu\nu}^{\bm{k}_{1}\bm{k}_2}(\bm{G})
        = \sum_{\bm{m}}^{N_{\tr{cell}}^{\tr{AFT}}} \me^{\mi\bm{k}_2\cdot\bm{m}}
        \int\md\bm{r}\, \phi_{\mu}(\bm{r}) \phi_{\nu}(\bm{r}-\bm{m})
        \me^{-\mi(\bm{k}_{12}+\bm{G})\cdot\bm{r}}
\end{equation}
is the AFT of the AO pair density.
Therefore, the cost of evaluating the LR 3c integrals also has two parts:
$O(N_k^2 N_{\tr{cell}}^{\tr{AFT}} N_{\tr{PW}} n_{\tr{AO}}^2)$ for the AFT of the AO pair density (\ref{eq:AFT_AOpair}) and $O(N_k^2 N_{\tr{PW}} n_{\tr{aux}} n_{\tr{AO}}^2)$ for the PW contraction in \cref{eq:LR3c}.

\section{Comparison of different strategies for batching the 3c integrals}
\label{sec:cmp_batch}

In principle, the 3c integrals (\ref{eq:3c_RSDF}) can be batched in three ways: by auxiliary indices, by AO indices, and by $k$-point indices.

For evaluating the SR 3c integrals, batching either the auxiliary or the AO indices introduces no extra cost, while batching the $k$-point indices requires repeated computation of the molecular integrals (\ref{eq:MolSR3c}).

For evaluating the LR 3c integrals, batching the auxiliary indices alone does not remove the storage bottleneck because the AFT of the AO pair densities (\ref{eq:AFT_AOpair}) needs $O(N_k^2 N_{\tr{PW}} n_{\tr{AO}}^2)$ storage, which is similar to the storage cost of the 3c integrals that we aim to avoid.
Batching the AO indices removes this storage bottleneck.
The potential repeated computation of the AFT of the auxiliary basis functions (\ref{eq:AFT_aux}) is unnecessary because $\tilde{\chi}_{P}^{\bm{k}_{12}}(\bm{G})$ needs only modest storage [$O(N_k N_{\tr{PW}} n_{\tr{aux}})$] and can readily be cached in memory.
Batching the $k$-point indices removes the storage bottleneck as well.
However, from \cref{eq:AFT_AOpair}, it is clear that the AFT of the AO pair density is most efficiently performed for a group of AO pair densities that have the same crystal momentum transfer, $\bm{k}_{12}$.
This limits the minimal batch size for the $k$-point index pairs $(\bm{k}_1,\bm{k}_2)$ to be at least $N_k$.

From the discussion above, it is clear that batching the auxiliary indices is not an option for our purpose. (It may work, however, if one only needs the SR 3c integrals, e.g.,~for some hybrid KS-DFT calculations \cite{Heyd03JCP}.)
We choose to batch the AO indices in this work because it introduces no extra computational cost for evaluating both the SR and the LR integrals.
This choice is particularly appropriate for systems with large unit cells that require only modest $k$-point sampling, which leaves little room for one to batch the $k$-point indices.

Although not explored in this work, we note that for systems with a small unit cell and a dense $k$-point mesh, batching the AO indices may not be possible because even a minimal batch in Eq.~(M13), $W_{P\mu [i]}^{\bm{k}_1\bm{k}_2}$ for $[i]$ a single orbital, may exceed the available storage.
This affects both the $K$-build for HF and the AO-to-CO integral transform for MP2.
In that case, batching the $k$-point indices is the only option.
The extra computational cost for repeatedly evaluating the molecular integrals (\ref{eq:MolSR3c}) will be amortized for large $N_k$ since contracting the molecular integrals with the phase factors (\ref{eq:SR3c}) has a much higher scaling with $N_k$.
Let us assess the future need for such $\bm{k}$-batched integral-direct algorithms.
For systems with a small unit cell and a high-quality basis set, the three-center integrals $V_{P\mu\nu}^{\bm{k}_1\bm{k}_2}$ can be stored in memory or on disk for systems with $N_k \approx 5^3$.
For insulators, this is commonly large enough to allow accurate extrapolation of the energy to the thermodynamic limit, but calculations requiring larger $k$-point meshes will require an integral-direct implementation along the lines described here.

\clearpage

\bibliography{refs_si}